\documentclass[prl, twocolumn, byrevtex]{revtex4}
\usepackage{graphicx}
\usepackage{dcolumn}
\usepackage{bm}
\newcommand{\inganas}{Ga$_{1-x}$In$_x$N$_y$As$_{1-y}$}
\begin{document}
\preprint{{\em Manuscript prepared for Applied Physics Letters}}
\title{Alloy disorder effects on the room temperature optical properties of \inganas quantum wells}
\author{Bhavtosh Bansal$^\ast$, Abdul Kadir, Arnab Bhattacharya and B. M. Arora}
\affiliation{Department of Condensed Matter Physics and Materials
Science, Tata Institute of Fundamental Research, 1 Homi Bhabha Road, Mumbai-400005, India}
\author{Rajaram Bhat}
\affiliation{Corning Inc., SP PR 02-3, Sullivan Park, Corning, NY 14831, USA}
\date{\today}
\begin{abstract}
{
The effect of alloy disorder on the optical density of states and the average room temperature carrier statistics
in \inganas\ quantum wells is discussed. A red shift between the peak of the room temperature photoluminescence and
the surface photovoltage spectra, that systematically increases with the nitrogen content within the quantum wells
is observed. The relationship between this Stokes' shift and the absorption linewidth in different samples suggests that
the photoexcited carriers undergo a continuous transition $-$ from being in quasi-thermal equilibrium with the lattice, to
being completely trapped by the quantum dot-like potential fluctuations $-$ as the nitrogen fraction in the alloy is
increased. The values of the 'electron temperature' inferred from the photoluminescence spectra are found to be
consistent with this interpretation.
}
\end{abstract}
\maketitle
\inganas\ dilute nitride alloys\cite{iee-issue, klar-review} provide an attractive possibility of designing
quantum wells (QWs) which are optically active in the 1.3 $\mu$m range, provide strong carrier confinement
and control of strain and band alignments, and can be grown on GaAs substrates. A number of experiments
on this system indicate that the very reasons for the large downward bowing of the optical energy gap, namely the small
size and the much larger electronegativity of the nitrogen atoms in comparison with the group-V host, are also
the ones responsible for creating a distribution of localized states and compositional fluctuations in a diluted nitride
sample. Such disordered QWs may behave like ensembles of quantum dots, capable of localizing charge carriers
in the low energy tail arising out of spatially varying disorder strength. Signatures of this carrier
localization appear, for example, in the anomalies in the temperature dependence of the photoluminescence(PL)
peaks and linewidths. Many previous PL\cite{klar-review, chen, pinault, sun, kudrawiec}, PL excitation\cite{sun},
photoreflectance\cite{klar-review, chen, kudrawiec_pss} and surface photovoltage(SPV)\cite{dumitras} studies have
reported such anomalous behavior, along with a  systematic and significant degradation of the transition linewidths and luminescence efficiency with
increasing nitrogen content in the samples and a Stokes-like shift\cite{klar-review, kudrawiec, sun} between the
maxima of the optical density of states(inferred from PL excitation or photoreflectance) and the (typically low
temperature) emission spectra.

In this letter, we have attempted to add to the earlier qualitative studies on electron localization
with a more quantitative analysis within the framework of different theories of
Stokes' shift in disordered QWs\cite{yang, alessi, polimeni, patane, gurioli, runge, eliseev}.
\begin{figure}[!b]
\begin{center}
\resizebox{!}{11.3cm} {\includegraphics{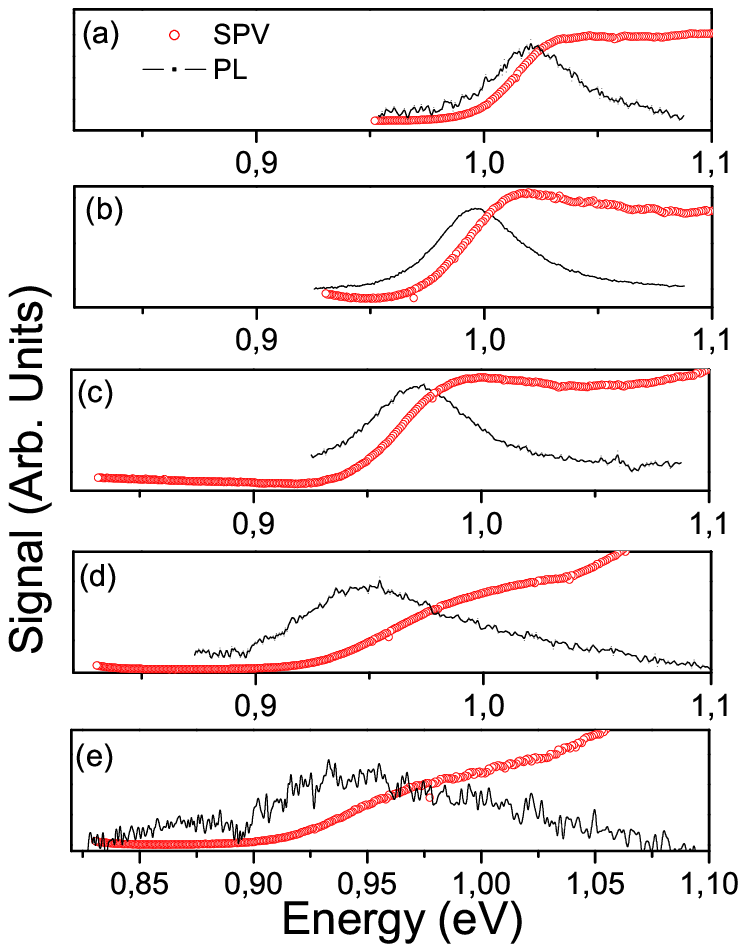}}
\caption{\label{fig:fig1}{\it SPV and PL spectra measured at room
temperature in \inganas\ QWs. See text for samples' details.}}
\end{center}
\end{figure}

\noindent {\em Experimental.} All the samples were grown by low
pressure metal-organic vapor phase epitaxy\cite{rbhat} on GaAs
substrates. The structure consisted of a GaAs buffer layer on
which a single $6-7$nm thick \inganas\ QW sandwiched within $\sim
8$nm GaAsN barrier layers was grown. This was capped by a 100nm
thick GaAs layer. The nitrogen content in the QW was varied from
$\sim 0.9\%$ to $1.7\%$ across the samples. The thickness and
composition of the QW and barrier for each of the samples was
determined by matching the high resolution X-ray diffraction
profile with that simulated using dynamical diffraction theory
({\em Philips X'Pert Epitaxy 3.0}). The values determined from the
simulation were consistent with the estimates made from the growth
parameters, as well as with  calculated PL transition energies
using the phenomenological model of Chow et al. \cite{chow}.

The disorder effects on optical properties we have studied here
are primarily determined by the changes in the nitrogen content of
the quaternary QW\cite{kent}. The small variations in nitrogen
content ($1.3-1.5\%$) and thickness($8-9$nm) in the GaAsN barrier,
as well as the variation in the QWs' indium content (40-44 \%, as
inferred from the X-ray simulations) have little effect on the
confinement energies and practically no effect on the linewidths
of interband transitions\cite{footnote2}. The samples in
Fig.\ref{fig:fig1} (a)-(e) have $y=0.009$, $0.011$, $0.013$,
$0.014$ and $0.017$ respectively with an approximate uncertainty
of around $\pm 0.001$ and $x=0.42\pm 0.02$.
\begin{figure}[!h]
\begin{center}
\resizebox{!}{7.6cm} {\includegraphics{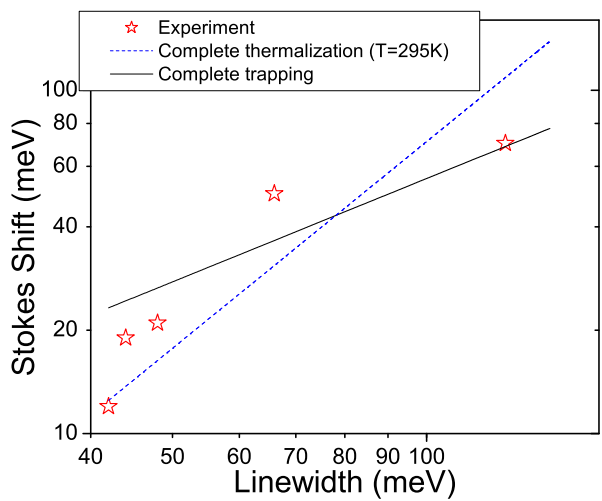}}
\caption{\label{fig:fig2}{\it(star) Approximate Stokes' shift(SS)
as a function of the SPV FWHM($\Delta$), inferred from the data in
Fig.\ref{fig:fig1}. (solid line) Prediction of the exciton
trapping\cite{yang} model, $\rm{SS}/\Delta$$=$$0.553$. (dotted
line) Prediction of the thermalization\cite{gurioli, eliseev}
model, $SS$$=$$0.18\Delta^2/k_BT_e$ with $T_e=295K$.}}
\end{center}
\end{figure}
SPV measurements were performed in `soft contact' mode. A detailed
discussion of the apparatus and the measurement technique has
already appeared in reference \cite{shouvik}. PL measurements were
made by non-resonant excitation of the samples at $488-514
\rm{nm}$ (Ar$^+$ laser) at a laser power density of approximately
$5$ watts per cm$^{-2}$. The collected signal was spectrally
dispersed using a monochromator. Both sets of measurements were
corrected for the system response. The values of Stokes' shift
were determined from the energy difference between the peak of the
PL spectrum and the shoulder in the SPV signal. The absorption
full width at half maxima (FWHM) were taken to be equal to twice
the energy difference between the mid-point and the shoulder of
the SPV spectrum. To determine the position of the SPV shoulder,
we utilized the distinct similarity between the overall shape of
the low energy part of the PL spectrum and the rising edge of the
SPV spectrum. The PL spectrum was shifted along the energy axis
till the low energy part of the PL and SPV spectra coincided; this
shift was taken as the value of the Stokes' shift.
\\
\noindent {\em Results and Discussion.} Fig.\ref{fig:fig1} shows
the SPV and PL spectra measured on five different QW samples whose
compositions were discussed earlier. It is evident that (1) the
linewidth of the emission and sharpness of the SPV spectra degrade
with a shift in the transition energies to lower values
(increasing nitrogen content), and (2) the Stokes'-like shift also
increases with the linewidth. Quantitative estimates of these
observations are given in Fig.\ref{fig:fig2} where this Stokes'
shift is plotted against the SPV FWHM. This Stokes'-like shift
arises because an absorption measurement like SPV (essentially)
reflects the optical density of states but the emission spectrum
also necessarily has to involve the distribution of carriers
occupying these states just before radiative recombination. A
non-resonant PL experiment, like the one reported here, involves
carrier excitation in the GaAs  cap region. These carriers then
migrate to the \inganas\ QW and may randomly diffuse into any of
the local potential minima. If the strength of the energy barriers
corresponding to the local minima trapping these carriers is
larger than the thermal energy, carriers will not be able to
thermalize within the finite recombination times. The PL spectra
would then preferentially reflect the density of states $\rho_0$
corresponding to the local energy minima (`minimum
distribution'\cite{yang, runge}), i.e.,  $PL(E) \sim \rho_0(E)$.
The absorption spectra, on the other hand, should not discriminate
between the troughs and crests of the potential energy surface.
This leads to a universal picture of Stokes' shift, first proposed
by Yang, Wilkinson, et al.\cite{yang}, which is expected to be
valid in the limit of very large disorder strength, low
temperature or both. This theory predicts that the Stokes'
shift(SS) should be related to the absorption FWHM ($\Delta$),
$\rm{SS}/\Delta$$=$$0.553$.  On the other hand, in the high
temperature limit where the fluctuations in the potential energy
landscape are small compared to the lattice temperature and the
carries are in quasi-equilibrium with the lattice, one would
expect the PL to be also affected by the thermal occupation
factor, i.e., $PL(E) \sim \rho_0(E) \exp(-E/k_BT_e)$. With the
assumption of a Gaussian density of states, it trivially
follows\cite{gurioli, eliseev} that $SS$$=$$0.18\Delta^2/k_BT_e$
in this case. The predictions of the two theories are also plotted
in Fig.\ref{fig:fig2}. We observe a crossover from the
thermalization induced Stokes' shift to a disorder induced Stokes'
shift as the compositional disorder is increased in the samples,
with the least and most disordered samples falling
on the first and the second curves respectively. 
\begin{figure}[!t]
\begin{center}
\resizebox{!}{7.8cm} {\includegraphics{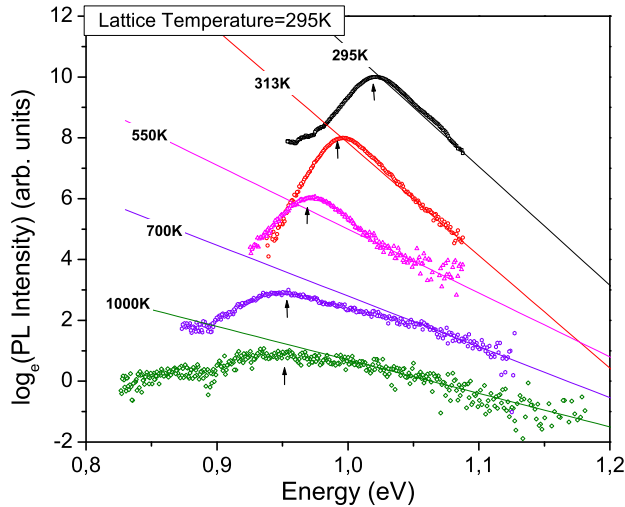}}
\caption{\label{fig:fig3}{\it The same room temperature PL spectra
as in Fig.\ref{fig:fig1}, but now plotted on a semi-log
scale(scatter plot). The straight lines (solid lines) depict
curves corresponding to different to the equation
$\log_{e}y=A-E/k_BT_e$ where $k_B$ is the Boltzmann constant, and
$T_e$ corresponds to an effective `electron temperature' marked on
each line and $A$ is an arbitrary number. Note that while the
samples with least amount of nitrogen has almost similar lattice
and electron temperatures, the effective electron temperature
increases as a function of nitrogen content within the QW (see
text).}}
\end{center}
\end{figure}

To gain further insight into the carrier distribution, the room
temperature PL plots are now depicted on a semilog scale in
Fig.\ref{fig:fig3}. The high energy tail of the spectrum is
expected to reflect the carrier distribution and can therefore be
fitted to the Boltzmann distribution function $A\exp(-E/k_BT_e)$,
where $k_B$ is the Boltzmann constant, $T_e$ is the `electron
temperature'\cite{gurioli} and $A$ is an undetermined constant
since the PL intensity is measured in arbitrary units. We observe
that this carrier temperature gets larger with the increase of the
nitrogen content in the samples. For the first sample, the carrier
temperature is almost exactly equal to the lattice temperature,
indicating almost complete thermalization, which was also the
conclusion of Fig.\ref{fig:fig2}, reached independent of this
analysis.

A simple interpretation of the high carrier temperature in
disordered samples follows from the observation that one can look
for an interpolation between the low and high temperature regimes
by combining the two previously discussed relations for the PL
spectra to a single expression of the kind, $PL(E) \sim \rho_0(E)
\exp(-\beta E/k_BT_e)$, with the parameter $0\leq \beta\leq 1$
continuously interpolating the carrier distribution between the
two extremes. Introduction of this parameter $\beta$
amounts\cite{bhavtosh_thermalization} to defining an effective
carrier temperature $T_e=T/\beta >T_{\rm{lattice}}$, which will be
larger in more disordered samples. The difference between carrier
and lattice temperature thus provides a convenient measure of the
sample quality.

\noindent {\em Conclusions.} We have presented a systematic
experimental study of alloying induced carrier localization at
room temperature in \inganas\ QWs. By independent assessments of
the (1) relationship between the Stokes' shift and absorption
linewidths, and (2) the carrier temperature inferred from the high
energy tail of the room temperature PL spectra in different
samples, we have been able to establish that there is a crossover
from the regime of complete carrier thermalization to the complete
localization regime as the alloying disorder is increased. These
methods provide quantitative means for a comparative assessment of
different \inganas\ samples.
\\
We thank Sandip Ghosh for many useful discussions and occasional
expert help during optical measurements and Mahesh Gokhale for
help in the XRD measurements.
\\
\noindent
{\footnotesize{$^*$Electronic Address: bhavtosh@tifr.res.in}}

\end{document}